\documentclass[12pt]{article}
\usepackage{amsmath}
\usepackage{amssymb}

\newtheorem{Theorem}{Theorem}[section]
\newtheorem{Proposition}[Theorem]{Proposition}
\newtheorem{Lemma}[Theorem]{Lemma}
\newtheorem{Definition}[Theorem]{Definition}
\newenvironment{Proof}[1]{{\bf Proof #1.} }{$\Box$\\}

\newcommand{\comment}[1]{}

\numberwithin{equation}{section}

\begin{document}

\centerline{\Large On dispersive effect of the Coriolis force}
\smallskip
\centerline{\Large for the stationary Navier-Stokes equations}

\bigskip

\smallskip

\centerline{Pawe\l{} Konieczny, Tsuyoshi Yoneda}

\begin{center}

{Institute for Mathematics and Its Applications}\\
{University of Minnesota}\\
{114 Lind Hall, 207 Church St. SE, Minneapolis, MN, 55455, U.S.A.}\\
{E-mail: \textsc{konieczny@ima.umn.edu}, \textsc{yoneda@ima.umn.edu}}
\end{center}

{\bf Abstract.} The dispersive effect of the Coriolis force for the stationary Navier-Stokes equations is investigated.
The effect is of a different nature than the one shown for the non-stationary case by 
J. Y. Chemin, B. Desjardins, I. Gallagher and E. Grenier.
Existence of a unique solution is shown for arbitrary large external force
provided the Coriolis force is large enough. The analysis is carried out in a new framework
of the Fourier-Besov spaces. In addition to the stationary case counterparts of several classical results
for the non-stationary Navier-Stokes problem have been proven. 

\medskip



\section{Introduction}

We consider the stationary 3D-Navier-Stokes equations with the Coriolis
force:

\begin{equation}\label{NSC}
(v \cdot \nabla) v+ \Omega e_3 \times v -\Delta v + \nabla p = F,\quad
 \nabla \cdot v=0\quad\mathrm{in~}\mathbb{R}^3,
\end{equation}
where $v=v(x)=(v^{1}(x),v^{2}(x), v^{3}(x))$ is the unknown
velocity vector field and $p=p(x)$ is the unknown scalar pressure
at the point $x=(x_1,x_2,x_3)\in \mathbb{R}^3$ in space, and $F$ is a given external force.
Here $\Omega \in \mathbb{R}$ is the Coriolis parameter, which is twice the angular
velocity of the rotation around the vertical unit vector
$e_3=(0,0,1)$, the kinematic viscosity coefficient is normalized by
one. By $\times$ we denote the exterior product, and hence, the
Coriolis term is represented by $e_3 \times u= Ju$ with the
corresponding skew-symmetric $3 \times 3$ matrix $J$.

Problems concerning large-scale atmospheric and oceanic flows are known
to be dominated by rotational effects. Almost all of the models of
oceanography and meteorology dealing with large-scale phenomena include
the Coriolis force. For example, oceanic circulation featuring a hurricane is caused by the large rotation.
There is no doubt that other physical effects are of similar significance like
salinity, natural boundary conditions and so on. However the first step
in the study of more complex model is to understand the behavior of
rotating fluids. To this end, we treat in a standard manner the Navier-
Stokes equations with the Coriolis force.

Let us look back the history of the Coriolis force. In 1868 Kelvin
observed
that a sphere moving along the axis of uniformly rotating water
takes with it a column of liquid as if this were a rigid mass (see
\cite{F} for references). After that, Hough \cite{H}, Taylor
\cite{Ta} and Proudmann \cite{P} made important contributions.
Mathematically it was investigated by Poincar\'e \cite{P}, more
recently, Babin, Mahalov and Nicolaenko \cite{BMN1,BMN2} considered non-stationary Navier-Stokes equations with Coriolis force 
in periodic case. The periodicity is extended to the almost periodic case
 by several authors.
For the results of local existence of non-stationary rotating Navier-
Stokes equations with spatially almost periodic data and its properties,
see \cite{GIMM2,GJMY,GMN}.
Moreover, for the results of global existence and long time existence
in the almost periodic setting, see \cite{GIMS,GIMS2, Y} for example.

 On the other hand, Chemin, Desjardins, Gallagher and Grenier
(CDGG)\cite{CDGG} considered decaying data case. 
CDGG derived dispersion estimates on a linearized version of the 3D-Navier-Stokes equations
with the Coriolis force to show existence of global solution to the non-stationary rotating Navier-Stokes system.
To construct such estimate, they handled eigenvalues and eigenfunctions
of the Coriolis operator.

The main result of this paper is to show existence of the solution to the stationary
Navier-Stokes equations with the Coriolis force for arbitrary large external force
provided that the Coriolis force is sufficiently large.
To do so, we handle new type of function spaces,
namely, Fourier Besov spaces (FB) which are designed to present in a clear way how the Coriolis force
has influence on the solution to the considered system.
A similar approach to introduce function spaces which make analysis of specific features of a system
much easier has been shown in a paper by the first author and P. B. Mucha in \cite{KM}, where they
investigate asymptotic structure of solution to the stationary Navier-Stokes equations in $\mathbb{R}^2$.

In FB spaces, we cannot expect to use energy type estimates and the
structure of Hilbert spaces as CDGG used.
The main motivation to introduce those spaces is that in this framework we are able
to present directly dispersive effect of the Coriolis force (see Proposition \ref{bigCoriolis}),
which is in principle different from the dispersive effect from CDGG.

To show usefulness of introduced spaces we prove existence to the non-stationary Navier-Stokes-Coriolis
system in function spaces which are counterparts for well known classical results in the Navier-Stokes
theory (see \cite{BS, CK, Cannone}).
Moreover we can considerably simplify other results for the Navier-Stokes-Coriolis system, like
recent results by Giga, Inui, Mahalov and Saal \cite{GIMS2}.

\subsection{Preliminaries}

In this section we would like to recall basic facts of Littlewood-Paley theory. We denote by $\varphi \in \mathcal{S}(\mathbb{R}^3)$ a radially
symmetric supported in $\{\xi \in \mathbb{R}^3 : \frac{3}{4} \leq |\xi| \leq \frac{8}{3}\}$ such that
$$
	\sum_{j\in\mathbb{Z}} \varphi(2^{-j}\xi) = 1\quad \textrm{for all~} \xi \neq 0.
$$
We also introduce the following functions:
$$\varphi_j(\xi) = \varphi(2^{-j}\xi),\quad \psi_j(\xi) = \sum_{k\leq j-1} \varphi_k(\xi).$$
Now we define standard localization operators:
\begin{equation}
 \Delta_j f = \varphi_j f,\qquad S_j f = \sum_{k\leq j-1}\Delta_k f = \psi_j f,\qquad \mathrm{for~} j\in \mathbb{Z}.
\end{equation}
It is then easy to verify the following identities:
\begin{eqnarray}
 \Delta_j \Delta_k f & = 0\qquad \mathrm{if~} |j-k| \geq 2,\\
 \Delta_j(S_{k-1} f \Delta_k f) & = 0 \qquad \mathrm{if~} |j-k| \geq 5.
\end{eqnarray}
Moreover one can follow Bony (see \cite{Bony}) and introduce the following decomposition:
\begin{equation}\label{fgDecomp}
	fg = T_f g + T_g f + R(f, g),
\end{equation}
where
\begin{equation}
 T_f g = \sum_{j\in\mathbb{Z}} S_{j-1} f \Delta_j g,\quad R(f, g) = \sum_{j\in \mathbb{Z}} \Delta_j f \tilde{\Delta_j}g, \quad
\tilde{\Delta_j} g = \sum_{|j' - j| \leq 1} \Delta_{j'} g.
\end{equation}

The framework for our results is determined by the Fourier-Besov spaces defined as follows:
\begin{Definition}
 We introduce the following homogeneous function spaces (called Fourier-Besov spaces):
	\begin{itemize}
	 \item $\dot{FB}^s_{p, q}(\mathbb{R}^n) = \{ f\in \mathcal{S}' : 
		\|f\|_{\dot{FB}^s_{p, q}(\mathbb{R}^n)} = \left(\sum_{k\in\mathbb{Z}} 2^{ksq}\|\varphi_k \hat{f}\|_{L_p(\mathbb{R}^n)}^q\right)^{1/q} < \infty \}$,
	 \item $\dot{FB}^s_{p, \infty}(\mathbb{R}^n) = \{ f\in \mathcal{S}' : 
		\|f\|_{\dot{FB}^s_{p, \infty}(\mathbb{R}^n)} = \sup_{k\in\mathbb{Z}} 2^{ks}\|\varphi_k \hat{f}\|_{L_p(\mathbb{R}^n)} < \infty \}$.
	\end{itemize}
\end{Definition}

In our considerations we are using results for the Stokes problem with the Coriolis force:
\begin{equation}
  \left\{ \begin{aligned}
		u_t - \nu\Delta u + \Omega e_3\times u + \nabla p & = F,\\
		\mathrm{div~} u & =  0,\\
		u(0, x) & = u_0(x).
	\end{aligned} \right.
\end{equation}
For this system one has the following formula for the solution (see \cite{GIMM2}):
\begin{equation}
 \hat{u}(t, \xi) = \cos\left(\Omega\frac{\xi_3}{|\xi|}t\right) e^{-\nu|\xi|^2 t} I \hat{u_0}(\xi) +
			\sin\left(\Omega\frac{\xi_3}{|\xi|}\right) e^{-\nu|\xi|^2 t} R(\xi) \hat{u_0}(\xi),
		\quad t \geq 0, \xi \in \mathbb{R}^3,
\end{equation}
where $I$ is the identity matrix and
\begin{equation}\label{RDef}
 	R(\xi) = \left(
		\begin{array}{ccc}
	          0 & \frac{\xi_3}{|\xi|} & -\frac{\xi_2}{|\xi|} \\
		 - \frac{\xi_3}{|\xi|} & 0 & \frac{\xi_1}{|\xi|}	\\
		\frac{\xi_2}{|\xi|} & -\frac{\xi_1}{|\xi|} & 0
	         \end{array}
		\right).
\end{equation}

An important observation is that
\begin{equation}
 |\hat{u}(t, \xi)| \leq e^{-\nu|\xi|^2 t}|\hat{u_0}(\xi)|,\quad t\geq 0, \xi\in\mathbb{R}^3.
\end{equation}

\section{Main results}

In this section we formulate our main results for the non-stationary and stationary Navier-Stokes equations with the Coriolis force.
We would like to mention that it is not difficult to obtain also other results (like stability of solutions to the non-stationary case) in this framework. We refer the Reader
to the paper by Cannone and Karch \cite{CK} as a reference for what can be expected. We do not prove those results to keep the paper more readable.

\subsection{Non-stationary case}

In the following theorem we consider mild solutions to the following non-stationary Navier-Stokes system with the Coriolis force:
\begin{eqnarray}
 u_t - \nu\Delta + \Omega e_3 \times u + u\cdot\nabla u + \nabla p & = & 0,\label{NSC1}\\
 \mathrm{div~} u & = & 0,\label{NSC2}\\
 u(0, x) & = & u_0(x).\label{NSC3}
\end{eqnarray}

\begin{Theorem}\label{NonstationaryMainTheorem}
 Let $\Omega \in \mathbb{R}$ be an arbitrary constant. Let $u_0 \in X_0$ and $\|u_0\|_{X_0}$ be small enough (independently of $\Omega$).
 Then there exists a unique global in time solution $u \in Y$ to problem (\ref{NSC1})-(\ref{NSC3}), where $X_0$ and $Y$ one can take as follows:
	\begin{itemize}
	 \item $X_0 = \dot{FB}^{2-3/p}_{p, \infty}$, $Y = C^w([0, \infty); \dot{FB}^{2-3/p}_{p, \infty}) \cap L^\infty(0, \infty; \dot{FB}^{2-3/p}_{p, \infty})$, where $3 < p \leq \infty$,
	 \item $X_0 = \dot{FB}^{2-3/p}_{p, p}$, $Y = C([0, \infty); \dot{FB}^{2-3/p}_{p, p})\cap L^\infty(0, \infty; \dot{FB}^{2-3/p}_{p, p})$, where $3 < p < \infty$,
	 \item $X_0 = \dot{FB}^{-1}_{1, 1}\cap \dot{FB}^0_{1, 1}$, $Y = C([0, \infty); \dot{FB}^{0}_{1, 1}) \cap L^2(0, \infty; \dot{FB}^{-1}_{1, 1}\cap \dot{FB}^{0}_{1, 1})$.
	\end{itemize}
 Moreover the following case is valid:
	\begin{itemize}
	 \item $X_0 = \dot{FB}^{2-3/p}_{p, \infty}$, \\
		$Y = L^\infty(0,\infty;\dot{FB}^{2-3/p}_{p, \infty}) \cap L^1(0,\infty;\dot{FB}^{4-3/p}_{p, \infty}) \cap C^w([0,\infty);\dot{FB}^{2-3/p}_{p, \infty})$, for $1 < p \leq \infty$,
	\end{itemize}
\end{Theorem}

{\textbf{Note:}} The mentioned cases have their counterparts in the current literature for the non-stationary Navier-Stokes equations.
For example the case $\dot{FB}^{2}_{\infty, \infty}$ was considered by Cannone and Karch in \cite{CK}, the case
$\dot{B}^{3/p-1}_{p,\infty}$ (which is a counterpart for $\dot{FB}^{2-3/p}_{p, \infty}$) in the paper \cite{Cannone} by Cannone. The case $\dot{FB}^{2-3/p}_{p, p}$ 
was treated by Biswas and Swanson for periodic case in \cite{BS}. Their result covers the whole range $1 < p \leq \infty$ due
to the periodicity -- more precisely in their case the authors do not have problems with integrability (summability) close to $0$ in the Fourier space.
In our case analysis close to $0$ in the Fourier space requires the assumption $p > 3$.
An analogue of the case $\dot{FB}^{-1}_{1, 1}\cap\dot{FB}^{0}_{1,1}$, that is $FM^{-1}_0 \cap FM_0$ spaces, has been published recently by Giga, Inui, Mahalov and Saal in \cite{GIMS2}.

In this paper consider those results in our setting, which seems to be more suitable for the Navier-Stokes equations
with the Coriolis force. Unfortunately using these methods we were not able to include the case $p = 2, q = 2$ which has been recently proven
by Hieber and Shibata in \cite{HS}.

\subsection{Stationary case}

In the following theorem we consider mild solutions to the Navier-Stokes system with the Coriolis force (\ref{NSC}).

\begin{Definition}
 For the sake of the stationary case with Coriolis force we introduce the following function space for $1 \leq p \leq \infty$,
	\begin{equation}\label{XOmega}
		X_{\mathcal{C}, \Omega}^p = \{ f\in \mathcal{S}' : \|f\|_{X_{\mathcal{C}, \Omega}^p} = \|w_1(\cdot)\hat{f}(\cdot)\|_{L^p} + \|w_2(\cdot)\hat{f}(\cdot)\|_{L^p} < \infty \},
	\end{equation}
	where
	\begin{equation*}
	 w_1(\xi) = \frac{|\xi|^{6-3/p}}{|\xi|^6 + \Omega^2|\xi_3|^2},\quad w_2(\xi) = \frac{\Omega|\xi_3||\xi|^{3-3/p}}{|\xi|^6 + \Omega^2|\xi_3|^2}R(\xi),
	\end{equation*}
	and $R(\xi)$ is the matrix (\ref{RDef}). 
\end{Definition}

The following theorem is the main result of our paper. 

\begin{Theorem}\label{StationaryMainTheorem}
 \textbf{Stationary case.} 
  Let $3 < p \leq \infty$. Then for all $F \in X_{\mathcal{C}, \Omega}^p$ such that 
  $\|F\|_{X_{\mathcal{C}, \Omega}^p}$ is small enough there exists a unique solution $u$ to the problem (\ref{NSC}) 
  such that $u\in \dot{FB}^{2-3/p}_{p,p}$ and the following estimate is valid:
\begin{equation}
 \|u\|_{\dot{FB}^{2-3/p}_{p,p}} \leq C\|F\|_{X_{\mathcal{C}, \Omega}^p}.
\end{equation}
 {\textbf{Note:}} Analogous result holds also for the spaces $\dot{FB}^{2-3/p}_{p, \infty}$.
\end{Theorem}

\textbf{Remark:}
 An important fact about the space $X_{\mathcal{C}, \Omega}^p$ is that $\dot{FB}^{-3/p}_{p, p} \subsetneq X_{\mathcal{C}, \Omega}^p$ for $\Omega \neq 0$ (see the proof of Proposition \ref{bigCoriolis}).
This means that the Coriolis force not only helps to weaken smallness assumptions on the force $F$ (see \cite{CK} and Lemma \ref{bigCoriolis} below)
but extends considerably the class of admissible external forces (for which we have existence result). For example the following function (it's Fourier transform):
\begin{equation}
\frac{|\xi|^6 + \Omega^2\xi_2^2}{|\xi|^6}\cdot 
 \left(
	\frac{\xi_1\xi_3}{|\xi|^2}, \frac{\xi_2\xi_3}{|\xi|^2}, \frac{\xi_3^2}{|\xi|^2}
 \right)
\end{equation}
is an element of $X_{\mathcal{C}, \Omega}^\infty$ for which (up to a constant) we have existence.
This function, however, is not an element of the space of pseudo-measures $\mathcal{PM} = \dot{FB}^0_{\infty, \infty}$ from the paper \cite{CK}.

In the case $F \in \dot{FB}^{-3/p}_{p, p}$ we can remove the smallness assumption provided that the Coriolis parameter $\Omega$
is large enough. This is being precised in the following Proposition (compare it with the case $F \in \dot{FB}^{-3/p}_{p, \infty}$
in Remark 2 after the proof of the Proposition).

\begin{Proposition}\label{bigCoriolis}
 	Let $3 < p < \infty$. Then for any given function $F \in \dot{FB}^{-3/p}_{p, p}$ there exists $\Omega_0$ such that 
	for all $\Omega \in \mathbb{R}$ satisfying $|\Omega| \geq \Omega_0$
	there exists the unique solution $u$ to problem (\ref{NSC}) such that $u\in \dot{FB}^{2-3/p}_{p,p}$.
\end{Proposition}

\begin{Proof}{of the Proposition}
 First we will show that for each $F \in \dot{FB}^{-3/p}_{p, p}$ and for all $\epsilon$ there exists $\Omega_0$ such that 
	for all $|\Omega| \geq \Omega_0$,
	\begin{equation}\label{FBXCIneq}
	 \|F\|_{X_{\mathcal{C}, \Omega}^p} \leq \epsilon\|F\|_{\dot{FB}^{-3/p}_{p, p}}.
	\end{equation}
 This fact together with Theorem \ref{StationaryMainTheorem} proves the Proposition.

 First we have $\dot{FB}^{-3/p}_{p, p} \subset X_{\mathcal{C}, \Omega}^p$. This is a simple observation since:
\begin{equation}\label{cosIneq}
 \frac{|\xi|^{4}}{|\xi|^6 + \Omega^2|\xi_3|^2} = \int_0^\infty e^{-t\xi^2} \cos(\Omega\frac{\xi_3}{|\xi|} t) dt \leq
	\int_0^\infty e^{-t|\xi|^2} dt= |\xi|^{-2}
\end{equation}
and
\begin{equation}\label{sinIneq}
 \frac{\Omega\xi_3|\xi|}{|\xi|^6 + \Omega^2|\xi_3|^2} = \int_0^\infty e^{-t\xi^2} \sin(\Omega\frac{\xi_3}{|\xi|} t) dt \leq
	\int_0^\infty e^{-t|\xi|^2} dt= |\xi|^{-2}.
\end{equation}

The proof of (\ref{FBXCIneq}) is fairly simple. First we decompose $\mathbb{R}^3$ into three regions:
	$\mathbb{R}^3 = A_{\delta} + B_{\delta} + C_{\delta}$, where
	$A_{\delta} = \{ \xi : |\xi_3| > \delta \wedge \delta < |\xi| < \frac{1}{\delta} \}$,
	$B_{\delta} = \{ \xi : |\xi_3| > \delta \wedge |\xi| > \frac{1}{\delta} \}$ and
	$C_{\delta} = \{ \xi : |\xi_3| < \delta\}$.

 For fixed $F$ there exists a compact set $K \subset \mathbb{R}^3$ such that $\||\xi|^{-3/p}F\|_{L^p(\mathbb{R}^3\setminus K)} \leq \epsilon/3$
 and by (\ref{cosIneq}) and (\ref{sinIneq}) we have the following estimates, uniform with respect to $\Omega$:
\begin{equation}
 \left(\frac{|\xi|^{6-3/p}}{|\xi|^6 + \Omega^2|\xi_3|^2}\hat{F}\right)^p \leq \left(|\xi|^{-3/p}\hat{F}\right)^p
\end{equation}
and
\begin{equation}
 \left(\frac{\Omega\xi_3|\xi|^{3-3/p}}{|\xi|^6 + \Omega^2|\xi_3|^2}R(\xi)\hat{F}\right)^p \leq \left(|\xi|^{-3/p}\hat{F}\right)^p.
\end{equation}

From the definition of $B_{\delta}$ and $C_{\delta}$ we get that $|K\cap(B_\delta\cup C_\delta)| \to 0$ as $\delta\to 0$, 
hence for $\delta$ small enough we have $\|F\|_{X_{\mathcal{C}, \Omega}^p(K\cap(B_\delta\cup C_\delta))} \leq \epsilon/2$.
Once $\delta$ is fixed we get back to the integral over $K\cap A_\delta$:
\begin{equation}
 \left(\int_{X_{\mathcal{C}, \Omega}^p(K\cap A_\delta)} \left(\frac{|\xi|^{6-3/p}}{|\xi|^6 + \Omega^2|\xi_3|^2}\hat{F}\right)^p d\xi\right)^{1/p}\leq
	\frac{(1/\delta)^6}{\delta^6 + \Omega^2 \delta^2} \|F\|_{\dot{FB}^{-3/p}_{p, p}} \leq \epsilon/4,
\end{equation}
 for $\Omega$ large enough (depending on $\epsilon$, $\delta$ and $\|F\|$).
Similarly
\begin{equation}
 \left(\int_{X_{\mathcal{C}, \Omega}^p(K\cap A_\delta)} \left(\frac{\Omega\xi_3|\xi|^{3-3/p}}{|\xi|^6 + \Omega^2|\xi_3|^2}\hat{F}\right)^p d\xi\right)^{1/p}\leq
	\frac{\Omega(1/\delta)^4}{\delta^6 + \Omega^2 \delta^2} \|F\|_{\dot{FB}^{-3/p}_{p, p}} \leq \epsilon/4,
\end{equation}
for $\Omega$ large enough.

This completes the proof.
\end{Proof}

{\textbf{Remark 2:}} The counterpart of Proposition \ref{bigCoriolis} for the case when $F\in \dot{FB}^{-3/p}_{p, \infty}$ 
requires additional assumptions on $F$. Method which we presented in the previous proof
requires smallness assumptions of the following form: there exists a number $K$ such that
\begin{equation}
 \sup_{|k| \geq K} 2^{-3k/p} \left(\|\varphi_k w_1 \hat F\|_{L^p} + \|\varphi_k w_2 \hat F\|_{L^p}\right) \quad \textrm{is small enough},
\end{equation}
where $w_1(\xi)$ and $w_2(\xi)$ are weights from the definition (\ref{XOmega}) of the space $X_{\mathcal{C}, \Omega}^p$.
In particular this condition allows one to have $\|F\|_{\dot{FB}^{-3/p}_{p, \infty}}$ arbitrary large not only in frequences
within the region $[-K, K]$ but also for $|k| \geq K$ provided weights $w_1$ and $w_2$ make them small enough.
The proof of this fact is analogous to the proof of Proposition \ref{bigCoriolis}.

\section{Proofs of main results}

\subsection{Proof of Theorem \ref{NonstationaryMainTheorem}}

We use a rather standard approach to show existence, namely via the following Banach fixed point theorem (\cite{CK}):
\begin{Lemma}
 Let $(\mathcal{X}, \|\cdot\|_{\mathcal{X}})$ be a Banach space and $B : \mathcal{X}\times \mathcal{X} \to \mathcal{X}$ a bounded
bilinear form satisfying $\|B(x_1, x_2)\|_{\mathcal{X}} \leq \eta \|x_1\|_{\mathcal{X}}\|x_2\|_{\mathcal{X}}$ for all $x_1, x_2 \in \mathcal{X}$
and a constant $\eta > 0$. Then if $0 < \epsilon < 1/(4\eta)$ and if $y\in \mathcal{X}$ such that $\|y\|_{\mathcal{X}} < \epsilon$, the equation
$x = y + B(x, x)$ has a solution in $\mathcal{X}$ such that $\|x\|_{\mathcal{X}} \leq 2\epsilon$. This solution is the only one in the ball
$\overline{B}(0, 2\epsilon)$. Moreover, the solution depends continuously on $y$ in the following sense: if $\|\tilde{y}\|_{\mathcal{X}} \leq \epsilon$,
$\tilde{x} = \tilde{y} + B(\tilde{x}, \tilde{x})$, and $\|\tilde{x}\|_{\mathcal{X}} \leq 2\epsilon$ then
$$
\|x-\tilde{x}\|_{\mathcal{X}} \leq \frac{1}{1-4\eta\epsilon} \|y-\tilde{y}\|_{\mathcal{X}}.
$$
\end{Lemma}

In our case the bilinear form $B$ is defined as follows:
\begin{equation}
 B(u, v)(t) = -\int_0^t \mathcal{G}(t-\tau)\mathbb{P}\mathrm{div}(u\otimes v)d\tau,
\end{equation}
where $\mathcal{G}$ was defined in (\ref{GDef}).

It is then straightforward that in order to prove existence we have to prove corresponding estimates in all cases of space $X$.

\begin{itemize}
 \item In case $X_0 = \dot{FB}^{2-3/p}_{p, \infty}$, where $3 < p \leq \infty$ we use Lemma \ref{GtConvF} with $r = \infty$ to get:
\begin{equation}
 \left\| \int_0^t \mathcal{G}(t-\tau)f(\tau)d\tau\right\|_{L^\infty_T(\dot{FB}^s_{p, \infty})} \leq \frac{1}{\nu} \|f\|_{L^\infty_T(\dot{FB}^{s-2}_{p, \infty})}.
\end{equation}
and then for $f = \mathrm{div}(u\otimes v)$ we use inequality (\ref{uvLinfty}). \\
Estimate for convolution with initial data $u_0$ comes from Lemma \ref{GtConv}.

\item In the case $X_0 = \dot{FB}^{2-3/p}_{p, p}$, where $3 < p < \infty$ we use Lemma \ref{GtConvPPtype} to estimate the bilinear form. Initial data $u_0$
estimates trivially. 

\item In the case $X_0 = \dot{FB}^{-1}_{1,1} \cap \dot{FB}^0_{1, 1}$ we make two steps. First we use Lemma \ref{GIMSGtConvF} with $s = 0$ 
combined with Lemma \ref{GIMSuv} (inequality (\ref{uvL1FB0})) to estimate bilinear form $B(u, v)$ in the space $L^2([0, \infty); \dot{FB}^{0}_{1,1})$ and
Lemma \ref{GIMSu0} with $s = 0$ to estimate initial data $u_0$. This gives us the unique solution in the space $L^2([0, \infty); \dot{FB}^{0}_{1,1})$.
In the second step we notice that using inequality (\ref{uvL1FB1}) and again Lemma \ref{GIMSGtConvF} with $s = 1$ we obtain that the solution is in fact in the space
$L^2([0, \infty); \dot{FB}^{1}_{1,1}\cap\dot{FB}^{0}_{1,1})$. This improved regularity is essential to show (in an elementary way) strong continuity of the solution, i.e.
$u \in C([0, \infty); \dot{FB}^0_{1,1})$.
\end{itemize}

To prove the second part of Theorem \ref{NonstationaryMainTheorem}, that is for $1 < p \leq \infty$ one uses the same results as in the case $3 < p \leq \infty$
but with estimate (\ref{uvL1}). Since this cases are of less interest to us (our paper focuses on the stationary case) we do not include more details
in order to keep the paper more consistent.

\subsection{Proof of Theorem \ref{StationaryMainTheorem}}

To prove existence results in the stationary case one may use the results from Theorem \ref{NonstationaryMainTheorem} in case $3 < p \leq \infty$ and
$X = \dot{FB}^{2-3/p}_{p, \infty}$ or $X = \dot{FB}^{2-3/p}_{p, p}$ and repeat reasoning from the paper by Cannone and Karch \cite{CK}.
The authors there use the following Lemma which is essential to obtain this result:
\begin{Proposition}\label{equiStat}
 The following two facts are equivalent
   \begin{itemize}
	\item $u = u(x)$ is a stationary mild solution to the problem (\ref{NSC1})-(\ref{NSC2}), that is
		\begin{equation}
		u = \mathcal{G}(t)u - \int_0^t \mathcal{G}(t-\tau)\mathbb{P}\mathrm{div~}(u\otimes u)d\tau + \int_0^t \mathcal{G}(\tau)\mathbb{P}F d\tau	
		\end{equation}
		for every $t > 0$.
	\item $u$ satisfies the integral equation
		\begin{equation}
			u = - \int_0^\infty \mathcal{G}(t-\tau)\mathbb{P}\mathrm{div~}(u\otimes u)d\tau + \int_0^\infty \mathcal{G}(\tau)\mathbb{P}F d\tau	
		\end{equation}
   \end{itemize}
\end{Proposition}

 Using this proposition and results for non-stationary case we see that in order to obtain existence of solution using a fixed point argument we just need to obtain estimates
 for the term with the force $F$. We use the formula for the Stokes-Coriolis semigroup, that is:
\begin{equation}\label{GDef}
 \hat{\mathcal{G}}(t) = \cos\left(\Omega\frac{\xi_3}{|\xi|}t\right) e^{-\nu|\xi|^2 t} I  +
			\sin\left(\Omega\frac{\xi_3}{|\xi|}\right) e^{-\nu|\xi|^2 t} R(\xi).
\end{equation}
Integrating this formula with respect to $t$ from $0$ to $\infty$ we get:
\begin{equation}
 \int_0^\infty \hat{\mathcal{G}}(t) dt = \frac{|\xi|^4}{|\xi|^6 + \xi_3^2 \Omega^2} I + \frac{\xi_3|\xi|}{|\xi|^6+\xi_3^2\Omega^2} R(\xi)
\end{equation}
It is then straightforward (from the definition of $\mathcal{X}_{\mathcal{C}, \Omega}^p$) that
\begin{equation}
 \left \|\int_0^\infty \mathcal{G} F dt\right\|_{\dot{FB}^{2-3/p}_{p, p}} \leq \|F\|_{\mathcal{X}_{\mathcal{C}, \Omega}^p}.
\end{equation}

\subsection{Main estimates}

\begin{Lemma}\label{basicIneq}
  For $1 \leq q \leq p \leq \infty$ and any multiindex $\gamma$ the following inequalities are valid:
	\begin{itemize}
	 \item $\mathrm{supp~}\hat{f} \subset \{|\xi| \leq A 2^j\} \Rightarrow 
			\|(i\xi)^\gamma \hat f\|_{L^q(\mathbb{R}^n)} \leq C 2^{j|\gamma|+nj(\frac{1}{q}-\frac{1}{p})} \|\hat f\|_{L^p(\mathcal{R}^n)}.$
	 \item $\mathrm{supp~}\hat{f} \subset \{ B_1 2^j \leq |\xi| \leq B_2 2^j\} \Rightarrow
			\|\hat f\|_{L^q(\mathbb{R}^n)} \leq C 2^{-j|\gamma|} \sup_{|\beta| = |\gamma|} \|(i\xi)^\beta \hat f\|_{L^p(\mathcal{R}^n)}.$
	\end{itemize}
\end{Lemma}

\begin{Lemma}\label{GtConv}
 	For $p\in [1, \infty]$ and $u_0 \in \dot{FB}^{2-3/p}_{p, \infty}$ one has: 
	\begin{equation}
	 \|\mathcal{G}(t) u_0 \|_{L_\infty(0, T; \dot{FB}^{2-3/p}_{p, \infty} \cap L_1(0, T; \dot{FB}^{4-3/p}_{p, \infty}))} \leq \max{}(1, \frac{1}{\nu})\|u_0\|_{\dot{FB}^{2-3/p}_{p, \infty}}.
	\end{equation}
	Moreover one also has:
	\begin{equation}
	 \|\mathcal{G}(t) u_0 \|_{L^\infty_T(\dot{FB}^s_{p, p})} \leq \|u_0\|_{\dot{FB}^{s}_{p, p}}.
	\end{equation}

\end{Lemma}

\begin{Proof}{\!}
	While the second estimate is straightforward let us focus on the first inequality.
	We consider the case $p < \infty$. The case $p = \infty$ can be obtained analogously.
	Let us first estimate the norm $\|\mathcal{G}(t) u \|_{L^\infty_T(\dot{FB}^{2-3/p}_{p, \infty})}$:
	\begin{equation*}
	 \|\mathcal{G}(t) u \|_{L^\infty_T(\dot{FB}^{2-3/p}_{p, \infty})} \leq \sup_{0\leq t < T} \sup_{k} 2^{k(2-3/p)} \|\varphi_k \hat{u_0} \|_{L^p} \leq \|u_0\|_{\dot{FB}^{2-3/p}_{p, \infty}}
	\end{equation*}
	The second part estimates as follows:
	\begin{multline}
	 \|\mathcal{G}(t) u \|_{L^1_T(\dot{FB}^{4-3/p}_{p, \infty})} \leq \int_0^T \sup_{k} 2^{k(4-3/p)}e^{-\nu t 2^{2k}} \|\varphi_k \hat {u_0}\|_{L^p} dt \\
		\leq \sup_{k} \frac{1}{\nu} 2^{-2k} 2^{k(4-3/p)} \|\varphi_k \hat{u_0} \|_{L^p} dt \leq \frac{1}{\nu}\|u_0\|_{\dot{FB}^{2-3/p}_{p, \infty}}.
	\end{multline}
	This finishes the proof of this Lemma.
\end{Proof}

\begin{Lemma}\label{GtConvF}
 	Let $s\in\mathbb{R}$, $p, q\in [1, \infty]$ and $f \in L^r_T(\dot{FB}^{s}_{p, \infty})$. Then the following estimate is valid:
	\begin{equation}
	\left \|\int_0^t \mathcal{G}(t-\tau)f(\tau)d\tau \right\|_{L^q_T(\dot{FB}^s_{p, \infty})} \leq \frac{1}{\nu}\|f\|_{L^r_T(\dot{FB}^{s-2-2/q+2/r}_{p, \infty})}.
	\end{equation}
\end{Lemma}

\begin{Proof}{\!}
	Since 
	\begin{equation*}
	 \left\|\int_0^t \mathcal{G}(t-\tau)f(\tau)d\tau \right\|_{L^q_T(\dot{FB}^s_{p, \infty})} = \sup_k 2^{sk} \left\|\int_0^t \|\hat{\mathcal{G}}(t-\tau)\hat{f}(\tau)\varphi_k\|_{L^p}d\tau \right\|_{L^q_T}
	\end{equation*}
	we may fix $k$ and estimate the corresponding term:
	\begin{eqnarray*}
	 2^{ks} \left\|\int_0^t \|\hat{\mathcal{G}}(t-\tau)\hat{f}(\tau)\varphi_k\|_{L^p}d\tau \right\|_{L^q_T}
		& \leq & 2^{ks}\left\|\int_0^t e^{(t-\tau)2^{2k}}\|\hat{f}(\tau)\varphi_k\|_{L^p}d\tau \right\|_{L^q_T}
	\end{eqnarray*}
	Using Young's inequality with $\tilde{q}$ such that $1 + \frac{1}{q} = \frac{1}{\tilde{q}} + \frac{1}{r}$, that is:
	$\frac{1}{\tilde{q}} = 1 + \frac{1}{q} - \frac{1}{r}$ we get:
	\begin{eqnarray*}
		2^{ks}\left\|\int_0^t e^{(t-\tau)2^{2k}}\|\hat{f}(\tau)\varphi_k\|_{L^p}d\tau \right\|_{L^q_T} \leq
			2^{ks}\|e^{t2^{2k}}\|_{L^{\tilde{q}}_T}\|\hat{f}(t)\varphi_k\|_{L^r_T(L^p)} \\ 
			\leq 2^{k(s-2 - \frac{2}{q} + \frac{2}{r})}\|\hat{f}(t)\varphi_k\|_{L^r_T(L^p)}.
	\end{eqnarray*}
	Taking supremum over all $k\in\mathbb{Z}$ one obtains the desired estimate.
\end{Proof}

\begin{Lemma}
	The following estimates are valid:
	\begin{itemize}
	 \item For $1 < p \leq \infty$ and $V = L^\infty_T(0, T; \dot{FB}^{2-3/p}_{p, \infty}(\mathbb{R}^3) \cap L^1_T(0, T; \dot{FB}^{4-3/p}_{p, \infty}))$ (introduced here for readability):
		\begin{equation}\label{uvL1}
			\|uv\|_{L^1_T(\dot{FB}^{3-\frac{3}{p}}_{p,\infty})} \leq \|u\|_{V}\|v\|_{V}.
		\end{equation}
	\item For $p > 3$:
		\begin{equation}\label{uvLinfty}
			\|uv\|_{L^\infty_T(\dot{FB}^{1-\frac{3}{p}}_{p,\infty})} \leq \|u\|_{L^\infty_T(\dot{FB}^{2-3/p}_{p, \infty})}\|v\|_{L^\infty_T(\dot{FB}^{2-3/p}_{p, \infty})}.
		\end{equation}
	\end{itemize}

\end{Lemma}
\begin{Proof}{\!}
In the following proof we follow in principle the reasoning from \cite{CMZ}.
 Let us focus on the first inequality. From the definition we have:
	\begin{equation}\label{LemBi10}
		\|uv\|_{L^1_T(\dot{FB}^{3-\frac{3}{p}}_{p,\infty})} = 
			\int_0^T \sup_{j} 2^{j(3-\frac{3}{p})} \|\Delta_j (uv)\|_{L_p} dt.
	\end{equation}
	For $\Delta_j(uv)$ we use decomposition (\ref{fgDecomp}), that is:
	\begin{equation}
	 \Delta_j(uv) = \sum_{|k-j|\leq 4} \Delta_j (S_{k-1} u\Delta_k v) +
			\sum_{|k-j|\leq 4} \Delta_j (S_{k-1} v\Delta_k u) +
			\sum_{k\geq j-2} \Delta_j (\Delta_k u \tilde{\Delta_k} v),
	\end{equation}
	and denote each corresponding integral from (\ref{LemBi10}) as $I_j$, $II_j$ and $III_j$.

	\begin{equation}
	 I_j = \int_0^T 2^{j(3-3/p)} \| \sum_{|k-j| \leq 4} \varphi_j (\psi_{k-1} \hat u \ast \varphi_k \hat v)\|_{L^p} dt 
		\leq \int_0^T 2^{j(3-3/p)} \sum_{|k-j| \leq 4} \|\psi_{k-1} \hat u\|_{L^1} \| \varphi_k \hat v\|_{L^p}dt.
	\end{equation}
	Now using Lemma \ref{basicIneq} we have the following inequality:
	\begin{equation}
	 \|\psi_{k-1}\hat u\|_{L^1} \leq \sum_{k' < k} \|\varphi_{k'}\hat u\|_{L^1} \leq \sum_{k' < k} 2^{k'3(1-1/p)} \|\varphi_{k'} \hat u\|_{L^p},
	\end{equation}
	which allows us to estimate $I_j$ as follows:
	\begin{eqnarray*}
	 I_j	& \leq & \int_0^T 2^{j(3-3/p)} \sum_{|k-j| \leq 4} \sum_{k' < k} 2^{k'} 2^{k'(2-3/p)}\|\varphi_{k'} \hat u\|_{L_p} \| \varphi_k \hat v\|_{L_p} dt\\
		& \leq & \int_0^T 2^{j(3-3/p)} \sum_{|k-j| \leq 4} 2^k \sup_{k'} 2^{k'(2-3/p)}\|\varphi_{k'} \hat u\|_{L_p} \| \varphi_k \hat v\|_{L_p}dt \\
		& \leq & \int_0^T 2^{j(4-3/p)} \| \varphi_k \hat v\|_{L_p} dt \sup_{k'}2^{k'(2-3/p)}\|\varphi_{k'} \hat u\|_{L_p} \\
		& \leq & \|v\|_{L^1_T(\dot{FB}^{4-3/p}_{p, \infty})} \|u\|_{L^\infty_T(\dot{FB}^{2-3/p}_{p, \infty})},
	\end{eqnarray*}
	where we used the fact that since $|j-k| < 4$ then $2^{j} \sim 2^{k}$.

	Integral $II_j$ is easily estimated in the same way as $I_j$. We will now focus on integral $III_j$.

	\begin{eqnarray*}
	 III_j	& = & \int_0^T 2^{j(3-3/p)} \sum_{k\geq j-2} \|\varphi_j \varphi_k u \tilde \varphi_k v\|_{L^p} 
		\leq \int_0^T 2^{j(3-3/p)}\sum_{k\geq j-2} \|\varphi_k \hat u \|_{L^1} \|\varphi_k \hat v\|_{L^{p}}\\
	 	& =    & \int_0^T \sum_{k\geq j-2} 2^{(j-k)(3-3/p)} \|\varphi_k \hat u\|_{L^p} 2^{k(2-3/p)} 
				\|\tilde \varphi_k \hat v\|_{L^p} 2^{k(4-3/p)}, \\
		& \leq & \sup_k \|\varphi_k \hat u\|_{L^p} 2^{k(2-3/p)} \int_0^T \sup_k \|\tilde \varphi_k \hat v\|_{L^p} 2^{k(4-3/p)} dt 
		\leq \|u\|_{L^\infty_T(\dot{FB}^{2-3/p}_{p, \infty})}\|v\|_{L^1_T(\dot{FB}^{4-3/p}_{p, \infty})},
	\end{eqnarray*}
where we again used Lemma \ref{basicIneq}.

In order to obtain estimate (\ref{uvLinfty}) one proceeds in a similar way as for the case of (\ref{uvL1}), applying proper changes like $3-3/p$ is replaced by $1-3/p$.
The requirement that $p > 3$ comes from estimate of $III_j$, that is in the case of (\ref{uvL1}) one has the term $\sum_{k\geq j-2} 2^{(j-k)(3-3/p)},$
which is finite for $p > 1$, while in case of estimate (\ref{uvLinfty}) one encounters the term $\sum_{k\geq j-2} 2^{(j-k)(1-3/p)},$ which is finite for $p > 3$.

\end{Proof}

In what follows we focus on estimates for the space $L^2_T(\dot{FB}^0_{1, 1})$.

\begin{Lemma}\label{GIMSu0}
  The following estimate is valid:
	\begin{equation}
	 \|e^{t\Delta} u_0\|_{L^2_T(\dot{FB}^s_{1,1})} \leq \|u_0\|_{\dot{FB}^{s-1}_{1,1}}.
	\end{equation}
\end{Lemma}

\begin{Proof}{}
 This inequality is easily obtained:
	\begin{multline}
	 \|e^{t\Delta} u_0\|_{L^2_T(\dot{FB}^s_{1,1})} = \left(\int_0^T \left(\sum_k \int_{\mathbb{R}^3} \varphi_k e^{-t\xi^2}|\xi|^{s} u_0(\xi) d\xi\right)^2d\tau \right )^{1/2} \\
		\leq \sum_k \left(\int_0^T e^{-t2^{2k+1}}2^{sk} \|\varphi_k u_0(\xi)\|_{L^1}^2 d\tau\right)^{1/2} \leq \sum_k 2^{(s-1)k} \|\varphi_k u_0(\xi)\|_{L^1} = \|u_0\|_{\dot{FB}^{s-1}_{1, 1}}
	\end{multline}
\end{Proof}

\begin{Lemma}\label{GIMSuv}
 The following estimate is valid:
	\begin{equation}\label{uvL1FB0}
	 \|uv\|_{L^1_T(\dot{FB}^0_{1, 1})} \leq \|u\|_{L^2_T(\dot{FB}^0_{1, 1})} \|v\|_{L^2_T(\dot{FB}^0_{1, 1})}.
	\end{equation}
 Moreover if $u, v\in L^2_T(\dot{FB}^0_{1, 1} \cap \dot{FB}^1_{1,1})$ then the following estimate is valid:
	\begin{equation}\label{uvL1FB1}
	 \|uv\|_{L^1_T(\dot{FB}^1_{1, 1})} \leq \|u\|_{L^2_T(\dot{FB}^0_{1, 1} \cap \dot{FB}^1_{1,1})} \|v\|_{L^2_T(\dot{FB}^0_{1, 1} \cap \dot{FB}^1_{1,1})}.
	\end{equation}

\end{Lemma}
\begin{Proof}{}
 First we note that $f\in \dot{FB}^0_{1, 1} \Leftrightarrow \hat f \in L^1$. 
 Then our inequality (\ref{uvL1FB0}) is proven in the following way:
	\begin{multline*}
	 \|uv\|_{L^1_T(\dot{FB}^0_{1,1})} = \int_0^T \|\hat{u}\ast \hat{v}\|_{L^1} \leq \int_0^T \|\hat{u}\|_{L^1}\|\hat{v}\|_{L^1} \\
		\leq \|\hat u\|_{L^2_T(L^1)}\|\hat v\|_{L^2_T(L^1)} = \|u\|_{L^2_T(\dot{FB}^0_{1,1})}\|v\|_{L^2_T(\dot{FB}^0_{1,1})}
	\end{multline*}
 To prove inequality (\ref{uvL1FB1}) we proceed in a similar way:
	\begin{multline*}
	 \|uv\|_{L^1_T(\dot{FB}^1_{1,1})} = \int_0^T \int_{\mathbb{R}^n} |\xi| \int_{\mathbb{R}^n} \hat{u}(\xi-\eta, \tau) \hat{v} (\eta, \tau) d\eta d\xi d\tau\leq \\
		\int_0^T \int_{\mathbb{R}^n} \int_{\mathbb{R}^n} (|\xi-\eta|+|\eta|)\hat{u}(\xi-\eta, \tau) \hat{v} (\eta, \tau) d\xi d\eta d\tau \\
		\leq \|\xi\hat u(\xi)\|_{L^2_T(L^1)}\|\hat v\|_{L^2_T(L^1)} + \|\hat u(\xi)\|_{L^2_T(L^1)}\|\eta \hat v(\eta)\|_{L^2_T(L^1)},
	\end{multline*}
	which finishes the proof of the Lemma \ref{GIMSuv}.
\end{Proof}

\begin{Lemma}\label{GIMSGtConvF}
 The following inequality is valid:
	\begin{equation}
		\left\| \int_{0}^t \mathcal{G}(t-\tau)f(\tau)d\tau \right\|_{L^2_T(\dot{FB}^s_{1, 1})} \leq
 			\frac{1}{\nu} \|f\|_{L^1_T(\dot{FB}^{s-1}_{1, 1})}.
	\end{equation}
\end{Lemma}
\begin{Proof}{}
 As previously we use triangle and Young's inequality to obtain:
	\begin{multline*}
		\left\| \int_{0}^t \mathcal{G}(t-\tau)f(\tau)d\tau \right\|_{L^2_T(\dot{FB}^s_{1, 1})} \leq
			\left\| \sum_k \int_{0}^t e^{(t-\tau)2^{2k}}2^{sk} \|\varphi_k f(\tau)\|_{L^1}d\tau \right\|_{L^2_T} \\
		\leq \sum_k \left\| e^{t2^{2k}}\right\|_{L^2_T}2^{sk} \|\varphi_k f(\tau)\|_{L^1_T(L^1)} = \sum_k 2^{(s-1)k} \|\varphi_k f(\tau)\|_{L^1_T(L^1)} = \|f\|_{L^1_T(\dot{FB}^{s-1}_{1,1})}.
	\end{multline*}
\end{Proof}

In what follows we focus on estimates for the space $L^\infty_T(\dot{FB}^{2-3/p}_{p, p})$, where $p > 3$. 
\begin{Lemma}\label{GtConvPPtype}
 The following estimate is valid:
	\begin{equation}
	 \left\|\int_0^t \mathcal{G}(t-\tau)\nabla(u\otimes v)d\tau\right\|_{L^\infty_T(\dot{FB}^{2-3/p}_{p, p})} \leq \|u\|_{L^\infty_T(\dot{FB}^{2-3/p}_{p, p})} \|v\|_{L^\infty_T(\dot{FB}^{2-3/p}_{p, p})}
	\end{equation}
\end{Lemma}

\begin{Proof}{}
 First let us estimate the convolution $\hat{u}\ast \hat{v}$. We do this as follows:
  \begin{eqnarray*}
	\hat u \ast \hat v (\xi) & = & \int_{\mathbb{R}^3} \frac{1}{|\eta|^{2-n/p}}\frac{1}{|\xi-\eta|^{2-n/p}} |\xi-\eta|^{2-n/p}\hat{v}(\xi-\eta) |\eta|^{2-n/p}\hat{u}(\eta)d\eta \\
		& \leq & \left(\int_{\mathbb{R}^3} \left(\frac{1}{|\eta|^{2-n/p}}\frac{1}{|\xi-\eta|^{2-n/p}}\right)^{p'}d\eta\right)^{1/p'} \\
		& & \cdot \left(\int_{\mathbb{R}^3}\left(|\xi-\eta|^{2-n/p}\hat{v}(\xi-\eta)|\eta|^{2-n/p}\hat{u}(\eta)\right)^pd\eta\right)^{1/p}.
  \end{eqnarray*}
  Now in order to estimate the convolution $\frac{1}{|\xi|^{(2-n/p)\tilde{p}}} \ast \frac{1}{|\xi|^{(2-n/p)p'}}$ we use the well known fact
	\begin{equation}\label{fourierFrac}
		\mathcal{F}(|\xi|^{-\alpha})(x) = |x|^{\alpha - n},	
	\end{equation}
 for $0 < \alpha < n$. Taking the Fourier transform of this convolution we get:
\begin{equation*}
 \mathcal{F}\left(\frac{1}{|\xi|^{(2-n/p)p'}} \ast \frac{1}{|\xi|^{(2-n/p)p'}}\right)(x) = |x|^{2[(2-n/p)p' - n]}.
\end{equation*}
Now using inverse Fourier transform we get:
\begin{equation*}
 \frac{1}{|\xi|^{(2-n/p)p'}} \ast \frac{1}{|\xi|^{(2-n/p)p'}} \sim |\xi|^{-2[(2-n/p)p' - n] - n} = |\xi|^{-2p'(2-n/p) + n}.
\end{equation*}
This formula holds for $p > 3$ (in dimension $3$) in order to satisfy (two times) condition for validity of (\ref{fourierFrac}).
We thus obtained the following formula:
\begin{equation}
 \left(\int_{\mathbb{R}^3} \left(\frac{1}{|\eta|^{2-n/p}}\frac{1}{|\xi-\eta|^{2-n/p}}\right)^{p'}d\eta\right)^{1/p'} \sim |\xi|^{-2(2-n/p)+n/p'}.
\end{equation}
 Going back to our main estimate:
   \begin{multline*}
	\left\|\int_0^t \mathcal{G}(t-\tau)\nabla(uv)d\tau\right\|_{L^\infty_T(\dot{FB}^{2-3/p}_{p, p})}  \leq \\
		\sup_t \left(\int_{\mathbb{R}^3} |\xi|^{(2-3/p)p} (\int_0^t e^{t\xi^2}dt)^p|\xi|^p |\hat{u}\ast\hat{v}(\xi)|^p d\xi\right)^{1/p} \\
		\leq \sup_t \left(\int_{\mathbb{R}^3} |\xi|^{A} |\xi|^{B}
			\left(\int_{\mathbb{R}^3}\left(|\eta|^{2-n/p}\hat{v}(\eta)|\xi-\eta|^{2-n/p}\hat{u}(\xi-\eta)\right)^pd\eta\right)^{p/p}d\xi\right)^{1/p},
   \end{multline*}
  where $A = 2p-3-2p+p$ and $B = [-2(2-n/p)+n/p']\cdot p$.

It is not hard to notice that $A + B = 0$ and thus the proof of the lemma follows easily from integration of the last term first with respect
to $\xi$ and then $\eta$.

\end{Proof}

\textbf{Acknowledgments: } Authors are grateful to Professor Vladimir {\v{S}}ver{\'a}k for his encouragement and guidance.
They also thank the Institute for Mathematics and Its Applications for support of their presence there during the academic year 2009/2010. 
The first author was partially supported by Polish grant No. N N201 547 438.

\end{document}